\documentclass[aps,prl,reprint]{revtex4-1}
\usepackage{graphicx}
\usepackage{url}
\usepackage{bm}
\usepackage{color}
\begin{document}

%\preprint{}

\title{Prediction of an Extended Ferroelectric Clathrate}


\author{Li Zhu} 
\email{z@zhuli.name}
\affiliation{Extreme Materials Initiative, Earth and Planets Laboratory, Carnegie Institution for Science, 5251 Broad Branch Road, NW, Washington, DC 20015, USA}

\author{Timothy A. Strobel}
\affiliation{Extreme Materials Initiative, Earth and Planets Laboratory, Carnegie Institution for Science, 5251 Broad Branch Road, NW, Washington, DC 20015, USA}

\author{R. E. Cohen}
\email{rcohen@carnegiescience.edu}
\affiliation{Extreme Materials Initiative, Earth and Planets Laboratory, Carnegie Institution for Science, 5251 Broad Branch Road, NW, Washington, DC 20015, USA}


\date{\today}

\begin{abstract}

Using first-principles calculations, we predict a lightweight room-temperature ferroelectric carbon-boron framework in a host/guest clathrate structure. This ferroelectric clathrate, with composition ScB$_3$C$_3$, exhibits high polarization density and low mass density compared with widely used commercial ferroelectrics.  Molecular dynamics simulations show spontaneous polarization with a moderate above-room-temperature T$_c$ of $\sim$370 K, which implies large susceptibility and possibly large electrocaloric and piezoelectric constants at room temperature. Our findings open the possibility for a new class of ferroelectric materials with potential across a broad range of applications. 

\end{abstract}

% \pacs{}
%\keywords{}

\maketitle


Ferroelectric materials combine a range of useful properties, placing them at the heart of many modern applications, such as non-volatile random access memory~\cite{Scott2007, Yang2008, Kim2009, Seidel2009}, actuators and sensors~\cite{Scott2007}, and electro-optical devices~\cite{Scott2007,Wessels2007}.  This variety of technological applications and the associated fundamental interest have motivated the design and development of new ferroelectric materials in the field~\cite{Caracas2007,Vadapoo2017,Bennett2013,Roy2012,Bennett2012,Garrity2013,Benedek2011,VanAken2004,Fennie2005,Benedek2014,Fennie2006,Oh2015,Lee2010,Choi2010,Ye2018,Liu2019}. New materials can give rise to improvements in physical properties and produce completely novel technological advancements. Since the first ferroelectric was discovered in Rochelle salt~\cite{Valasek1921}, the ferroelectric effect has been demonstrated in a broad range of materials including water-soluble crystals, oxides, polymers, ceramics, liquid crystals, and even molecular clathrates~\cite{Murakami1990,Fu2013,Zhang2014f,Tang2015}. Nevertheless, most of the commercial ferroelectric devices are based on the perovskite-type oxides of ABO$_3$. New types of ferroelectric materials can offer many advantages for various applications, such as small band gap ferroelectrics for solar energy conversion~\cite{Gou2011,Bennett2008}, and Pb-free ferroelectrics to reduce toxicity~\cite{Caracas2007,Vadapoo2017,Bennett2011,Ye2018}. The recently discovered hyperferroelectics were also reported to have a variety of exciting and potentially useful properties~\cite{Bennett2012,Garrity2013}. 
  
Although considerable progress has been made in the field of ferroelectrics, examples of  lightweight ferroelectrics with high strength are lacking. Many ferroelectrics are brittle, and their low toughness makes them susceptible to cracking under mechanical or electromechanical loads. Thus, the rational design of new ferroelectric materials with excellent mechanical properties is crucial to develop high performance intelligent devices. Recently, we predicted and synthesized a stable lightweight carbon-boron clathrate SrB$_3$C$_3$ (type-VII clathrate, $Pm\bar 3n$)~\cite{Zhu2020}. The SrB$_3$C$_3$ clathrate takes on the bipartite sodalite structure, which is composed of a strong $sp^3$-hybridized B-C framework with a high bulk modulus (experimental B$_0$ = 249(3) GPa). The prediction and synthesis of the first carbon-based clathrate creates opportunities for new lightweight materials with exceptional properties, such as those with high strength and tunable electronic structures by substituting different guest atoms. Given the observation of off-center cage occupancy in other clathrates~\cite{Sales2001, Christensen2016}, these new B-C frameworks present new opportunities for lightweight and robust ferroelectrics. However, all the previous off-centered extended clathrates are metallic, and none of them are ferroelectric.

This Letter proposes a strategy for designing lightweight clathrate ferroelectrics by introducing small cations into non-centrosymmetric variants of the newly discovered boron-carbon clathrate (SrB$_3$C$_3$). The clathrate cages are comprised of 24 vertices with alternating B and C atoms, and each cage contains a single guest Sr atom at the center. Small guest cations introduce unstable transverse optic phonon modes that tend to stabilize lower-energy polar structures, and the use of trivalent cations creates a condition of overall charge balance. We predict a scandium-filled boron-carbon clathrate, ScB$_3$C$_3$, as a model system for a new class of lightweight ferroelectrics. 

\begin{figure*}
    \includegraphics[width=17cm]{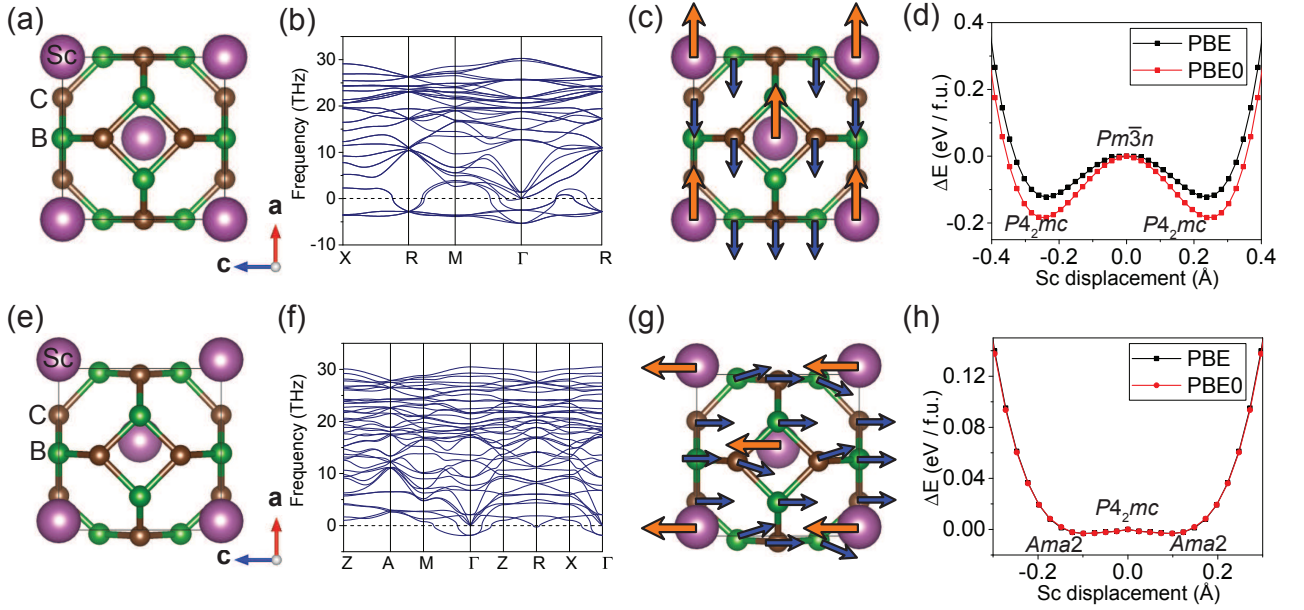}
    \caption{\label{fig1} (a) The crystal structure of ScB$_3$C$_3$ in the $Pm\bar 3n$ phase. (b) Phonon dispersion of the $Pm\bar 3n$ structure of ScB$_3$C$_3$. (c) Arrows indicate the eigenvector of the softened phonon mode at the $\Gamma$  point of the $Pm\bar 3n$ structure. (d) Total energies using PBE and PBE0 functionals for the $Pm\bar 3n$ structure for atoms displaced along the eigenvector of the transverse optical mode at the $\Gamma$ point. (e) The crystal structure of ScB$_3$C$_3$ in the $P4_2mc$ structure. (f) Phonon dispersion of the $P4_2mc$  structure of ScB$_3$C$_3$. (g) Arrows indicate the eigenvector of the softened phonon mode at the $\Gamma$  point of the $P4_2mc$ structure. (h) Total energies using PBE and PBE0 functionals for the $P4_2mc$  structure for the atoms displaced along the eigenvector of the transverse optical mode at the $\Gamma$ point. The purple, green, and brown spheres represent Sc, B, and C atoms, respectively.} 
    \end{figure*} 

Our first-principles calculations were mainly carried out in the framework of density functional theory (DFT) within the Perdew-Burke-Ernzerhof (PBE)~\cite{Perdew1996} and PBE0~\cite{Perdew1996a} functionals, as implemented in the Vienna ab initio Simulation Package (VASP) code~\cite{Kresse1996}. Our previous study verified that the PBE functional can reproduce the experimental results very well for the carbon-boron clathrate~\cite{Zhu2020}. However, the standard PBE approach typically underestimates the band gap of semiconductors. To obtain a more accurate electronic band structure, we adopted the PBE0 hybrid functional that contains 25\% of the exact Hartree-Fock exchange, 75\% of the PBE exchange, and 100\% of the PBE correlation energy ~\cite{Perdew1996a}. The PBE0 functional consistently yields more accurate band gaps for semiconductors ~\cite{Adamo1999}. The all-electron projector augmented wave (PAW) method~\cite{Blochl1994} was adopted with the pseudopotentials taken from the VASP library where $3s^23p^63d^14s^2$, $2s^22p^1$, and $2s^22p^2$ were treated as valence electrons for Sc, B, and C, respectively. The electronic wave functions were expanded in a plane-wave basis set with a cutoff energy of 520 eV. Monkhorst-Pack $k$-point meshes~\cite{Monkhorst1976} with a grid of spacing 0.04 $\times$ 2$\pi$ {\AA}$^{-1}$ for Brillouin zone sampling were chosen. To determine the dynamical stability of the studied structures, we performed phonon calculations by using the finite displacement approach, as implemented in the Phonopy code~\cite{Togo2015a}.

\begin{figure}
    \includegraphics[width=8.5cm]{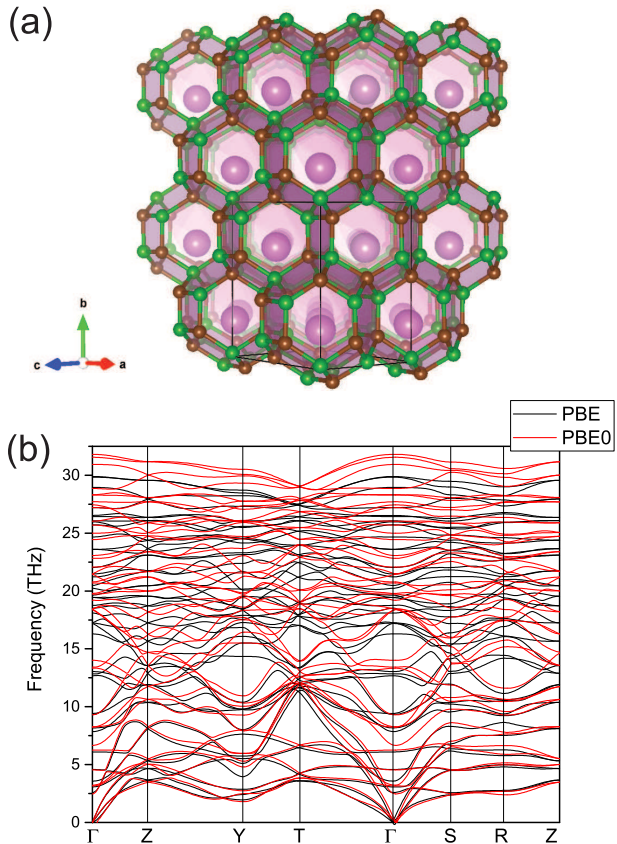}
    \caption{\label{fig2} (a) The predicted $Ama2$ polar structure of  ScB$_3$C$_3$ clathrate. The black box denotes the unit cell. The purple, green, and brown spheres represent Sc, B, and C atoms, respectively. (b) Phonon dispersion relations of the $Ama2$ structure of ScB$_3$C$_3$ calculated with PBE (black lines) and PBE0 (red lines) functionals, respectively. } 
\end{figure} 

The recently discovered SrB$_3$C$_3$ clathrate is metallic because Sr$^{2+}$[B$_3$C$_3$]$^{3-}$ is one electron per formula unit (f.u.) short of the octet rule~\cite{Zhu2020}. A ferroelectric must be an insulator so that it has a finite polarization that can be switched by an applied electric field. Based on the crystal structure of the SrB$_3$C$_3$  clathrate, we built an insulating ScB$_3$C$_3$ clathrate by substituting Sr$^{2+}$ with Sc$^{3+}$ (Fig. \ref{fig1}a). We computed the  electronic band structure of the $Pm\bar 3n$ structure of ScB$_3$C$_3$ using the PBE0 exact exchange-correlation functional (Fig. S1), and find  a band gap of 0.8 eV. We computed the phonon dispersion curves and examined the dynamic stability of the $Pm\bar 3n$ structure of ScB$_3$C$_3$ (Fig. \ref{fig1}b). We find unstable transverse optical phonons at the $\Gamma$ point, indicating structural instability. An imaginary phonon frequency (plotted as negative) corresponds to a saddle point on the multidimensional energy surface, and atomic displacements along this vibrational mode lead to an energy minimum or saddle point. The eigenvector of the most unstable phonon mode at the $\Gamma$ point (T$_{1u}$ mode, -5.09 THz) involves displacement of Sc atoms along the $a$ axis (Fig. \ref{fig1}c)). A double well is clear in the total energy versus normal mode coordinate (Fig. \ref{fig1}d) at the $\Gamma$ point for the cubic $Pm\bar 3n$ phase. The energy minima give a structure with space group $P4_2mc$ (Fig. \ref{fig1}e). We computed the phonon dispersions for this $P4_2mc$ structure, and found imaginary phonon frequencies at the $\Gamma$ point (Fig. \ref{fig1}f), indicating that the apparent minima in Fig. \ref{fig1}d are saddle points. This unstable phonon mode (E mode,  -2.02 THz) involves displacement of the Sc atoms along $c$ axis, and leads to the formation of an orthorhombic $Ama2$ phase. This structure is predicted to have all stable phonon modes (Fig. \ref{fig2}b). This newly predicted lower-symmetry $Ama2$ structure contains elongated cages in which the equilibrium guest Sc positions are located away from the cage center, resulting in a polar phase (Fig. \ref{fig2}a, see Supplemental Table 1 for structural parameters~\cite{supp}). Note that several unstable phonon modes appear throughout the Brillouin zone of the original cubic structure, which may lead to different minima. Thus, we also deformed the cubic $Pm\bar 3n$ phase along the combined eigenvectors of the different imaginary phonon modes, and the results show that the orthorhombic $Ama2$ phase remains the most stable structure (Fig. S2).

The $Ama2$ structure is composed of a single truncated octahedral cage with six four-sided faces and eight six-sided faces ($4^66^8$), which is similar to the structure of cubic SrB$_3$C$_3$ clathrate~\cite{Zhu2020}. The truncated octahedral cage is distorted from the perfect $O_h^3$ symmetry, resulting in four different bond distances between B and C atoms in the cage rather than one unique B-C bond length. This cage distortion is a result of  the Coulomb repulsive interaction between the off-centered Sc$^{3+}$ cation and the negatively charged cage, and is the driving force to produce the non-centrosymmetric structures.

\begin{figure}
    \includegraphics[width=8.5cm]{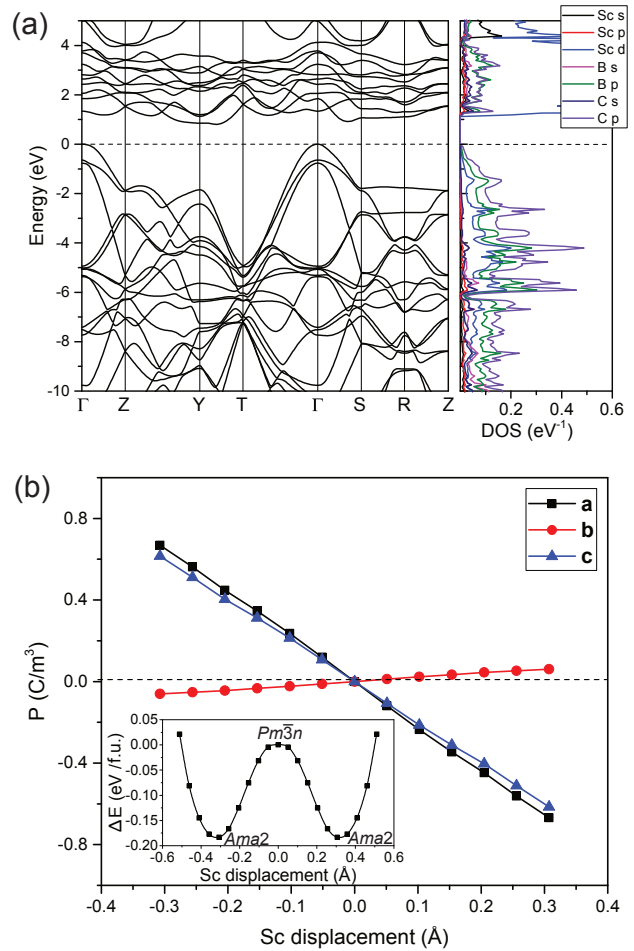}
    \caption{\label{fig3} (a) Electronic band structure and projected density of states for the $Ama2$ structure of ScB$_3$C$_3$ calculated with PBE0 functional. The Fermi energy is set to 0 eV (dashed line). (b) Calculated spontaneous polarization as a function of displacement along with the three parallel directions over the unit cell. The polar distortion is along the whole ferroelectric eigenvector that includes the displacements of Sc, B, and C atoms. The insert shows the characteristic ferroelectric double-well potential for the ScB$_3$C$_3$ clathrate. } 
\end{figure}

We computed the electronic band structure and the projected density of states (DOS) of the predicted ground state $Ama2$ structure (Fig. \ref{fig3}a). The top of the valence band is located at the $\Gamma$  point, and the  conduction band minimum is located at the T point ($\frac{1}{2}$,$\frac{1}{2}$,$\frac{1}{2}$) in the first Brillouin zone, which is consistent with SrB$_3$C$_3$ clathrate. The trivalent Sc atoms contribute three electrons to the clathrate framework resulting in the insulating phase, and the off-centered Sc atoms do not significantly change the band topology of the framework. The calculated projected density of states shows substantial overlap of the B-p and C-p bands, indicating strong covalent B-C bonding in the $Ama2$ structure. With PBE0, this structure is insulating, with an indirect band gap of $\sim$0.8 eV.

In the modern theory of polarization, the polarization is computed from the phase of the wave function using the Berry's phase approach~\cite{King-Smith1993}. To avoid ambiguity, we compute the polarization  for a path from the centrosymmetric structure (Fig. \ref{fig3}b) to the polar structure.
Fig. \ref{fig3}b shows the dipole moment as a function of polar distortion obtained by linear interpolation between the polar $Ama2$ and nonpolar $Pm \bar 3 n$ structures. 
The polarization along a and c are nearly equal, and the component along b is small. 
We obtain a polarization of 0.90 C/m$^2$ for the $Ama2$ phase. This is large compared with other ferroelectrics (e.g. 0.88 C/m$^2$ in PbTiO$_3$~\cite{Saghi-Szabo1998} and 0.7 C/m$^2$ in Pb(Mg,Nb)O$_3$~\cite{Kutnjak2006}). We obtain a similar polarization (0.85 C/m$^2$) with PBE in Quantum-Espresso ~\cite{Giannozzi2009} if we occupy states as if it were an insulator using the “fixed” occupation keyword. It is interesting that merely occupying the metallic states from PBE as if it were an insulator gives the same polarization as we find from the actual insulating state in PBE0.

 The energy difference between the polar state and the nonpolar, high-symmetry state is calculated to be $\sim$0.19 eV with PBE0 (Fig. \ref{fig3}b), which is in the range favorable for ferroelectric switching~\cite{PhysRevB.74.224412}. From the electron localization function (Fig. S4), we find strong covalent bonding between the B and C atoms, but only a weak interaction between Sc and the B/C atoms. These weak interactions allow the Sc atoms to switch between the two positions easily. We also find that the antiferroelectric phase ($Pmnn$, Fig. S5) is higher in energy than the ferroelectric phase, but the energy difference is only $\sim$28 meV per formula unit.  

\begin{figure}
    \includegraphics[width=7.5cm]{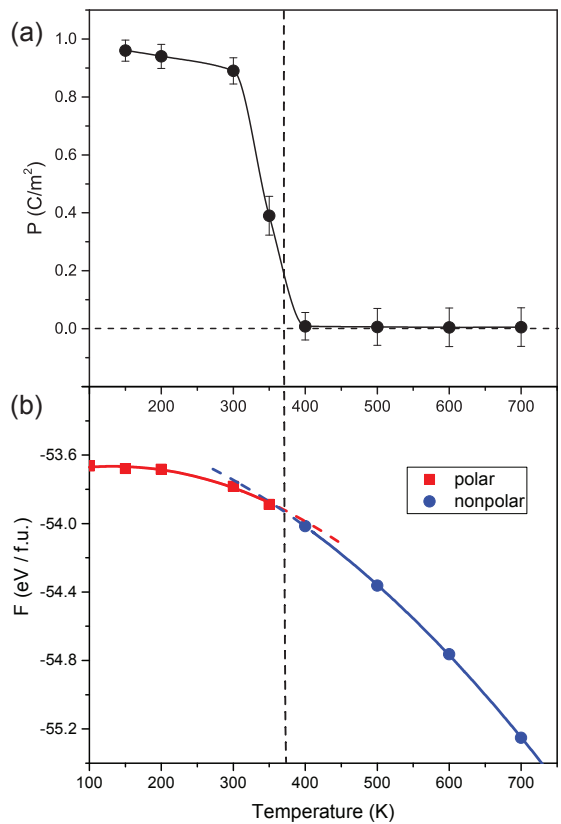}
    \caption{\label{fig4} (a) Spontaneous polarization of ScB$_3$C$_3$ clathrate as functions of temperature. (b) Helmholtz free energy of the polar and nonpolar phases as a function of temperature for ScB$_3$C$_3$. Short dash lines are the metastable extensions extrapolated from each phase.} 
\end{figure} 

An above-room-temperature Curie temperature (T$_c$) is desirable for  room-temperature operating conditions~\cite{Scott2007,Fu2013,Li2019,Huang2020}. We performed $NPT$ MD simulations using a 2 $\times$ 2 $\times$ 2 supercell (112 atoms) with temperature and pressure controlled via the Nose-Hoover thermostat and Parrinello-Rahman barostat, respectively. Each simulation was performed for 40 ps with a 1 fs time step. The average polarization $\bm{P}$ is calculated as $\bm{P} = \frac{1}{V}(\bm{Z}^*_{\text{Sc}}\bm{r}_{\text{Sc}} + 3\bm{Z}^*_{\text{B}}\bm{r}_{\text{B}} + 3\bm{Z}^*_{\text{C}}\bm{r}_{\text{C}})$, where $V$ is the volume of the primitive cell, $\bm{Z}^*_{\text{Sc}}$, $\bm{Z}^*_{\text{B}}$, and $\bm{Z}^*_{\text{C}}$ are the Born effective charges we compute for the Sc, B, and C atoms~\cite{supp}, and $\bm{r}_{\text{Sc}}$, $\bm{r}_{\text{B}}$, and $\bm{r}_{\text{C}}$ are the average displacements of the Sc, B, and C atoms. The evolution of the average polarization clearly shows a ferroelectric to paraelectric phase transition at $\sim$370 K (Fig. 4a), indicating it is ferroelectric at room temperature. 
To further investigate the ferroelectric to paraelectric phase transition, we compare their Helmholtz free energies. In this study, Gibbs free energy reduces to Helmholtz free energy at zero pressure. The Helmholtz free energy is computed from MD simulations as follows: $F=\langle U \rangle - \langle T \rangle S$, 
where $\langle\rangle$ represents ensemble average, $\langle U \rangle = \langle E \rangle + \langle E_{\text{kin}} \rangle$ is the sum of the Kohn-Sham energy and the kinetic energy of the ions, and the entropy $S$ is obtained by integrating the temperature-depended phonon frequencies that are the Fourier transform of the wave-vector dependent velocity autocorrelation function. The ferroelectric to paraelectric phase transition is also confirmed by Helmholtz free energy calculations (Fig. 4b). A T$_c$ slightly above room temperature bodes well for high response in electromechanical and electrocaloric response at room temperature. Thus, the clathrate may be suitable for many applications, such as temperature sensing, data storage, mechanical actuation, and energy harvesting~\cite{Scott2007,Fu2013}. 

At atmospheric pressure, the formation enthalpy of the $Ama2$ ScB$_3$C$_3$ structure is exothermic ($\Delta H = -163$ meV per formula unit). While we have not calculated the entire computationally demanding convex hull for this ternary system, the $Ama2$ structure of ScB$_3$C$_3$ is at least thermodynamically metastable, and the negative formation energy at atmospheric pressure is promising for future synthesis efforts.

In summary, we have used first-principles methods to predict ScB$_3$C$_3$ clathrate as a new type of ferroelectric material. The high spontaneous polarization makes ScB$_3$C$_3$ a promising material for piezoelectric and ferroelectric applications. Furthermore, the near room temperature T$_c$ implies large susceptibility and possibly large electrocaloric and piezoelectric constants at room temperature. The prediction of the first carbon-boron clathrate ferroelectric suggests the potential to access a wide range of previously undiscovered and inaccessible metastable ferroelectric materials. This work also provides a platform for investigating other types of clathrate structures through different elemental substitutions. There are prospects to obtain a much richer landscape for new ferroelectric materials with advanced properties by applying this design principle.  

This work is supported by U. S. Office of Naval Research Grants No. N00014-17-1-2768 and N00014-20-1-2699, and the Carnegie Institution for Science. Computations were supported by DOD HPC and Carnegie computational resources, and REC gratefully acknowledges the Gauss Centre for Supercomputing e.V. (www.gausscentre.eu) for funding this project by providing computing time on the GCS Supercomputer SuperMUC at Leibniz Supercomputing Centre (LRZ, www.lrz.de).

%\begin{acknowledgments}
%\end{acknowledgments} 

\bibliography{ScB3C3REF}

\end{document}